\documentclass[twocolumn,showpacs,amsmath,amssymb,superscriptaddress,prb]{revtex4}
\usepackage{amsmath,amssymb}
\usepackage{graphicx}
\usepackage{epstopdf}
\usepackage{color}
\usepackage{braket}

\begin{document}
\title{Quantum collapses and revivals of a matter wave in the dynamics of  symmetry breaking.}

\author{Frank Kirtschig}
\affiliation{Institute for Theoretical Solid State Physics, IFW Dresden, PF 270116, 01171 Dresden, Germany}
\affiliation{Department of Physics, Technical University Dresden, D-1062 Dresden, Germany}
\author{Jorrit Rijnbeek} 
\affiliation{Institute-Lorentz for Theoretical Physics, Universiteit Leiden, NL-2300 RA Leiden, The Netherlands}
\author{Jeroen van den Brink}
\affiliation{Institute for Theoretical Solid State Physics, IFW Dresden, PF 270116, 01171 Dresden, Germany}
\affiliation{Department of Physics, Technical University Dresden, D-1062 Dresden, Germany}
\author{Carmine Ortix}
\affiliation{Institute for Theoretical Solid State Physics, IFW Dresden, PF 270116, 01171 Dresden, Germany}

\pacs{64.60.Ht, 05.30.-d,05.70.Ln}

\date{\today}

\begin{abstract}
Using the remarkable mathematical construct of Eugene Wigner to visualize quantum trajectories in phase space, 
quantum processes can be described in terms of a quasi-probability distribution analogous to the phase space probability distribution of the classical realm.  
In contrast to the incomplete glimpse of the wave function that is achievable in a single shot experiment, the Wigner distribution, accessible by quantum state tomography, 
reflects the full quantum state. 
We show that during the fundamental symmetry-breaking process of a generic quantum system -- 
with a symmetry breaking field driving the quantum system far from equilibrium --  the Wigner distribution evolves continuously 
with 
 the system undergoing 
 a sequence of  revivals into the symmetry unbroken state, followed by collapses onto a quasi-classical state akin the one realised in infinite size systems.  
 We show that generically this state is completely delocalised both in momentum and in real space.   
 \end{abstract}

\maketitle

\section{Introduction}
\label{sec:intro}
Spontaneous symmetry breaking causes a macroscopic body under equilibrium condition  to have less symmetry than its microscopic building blocks. Probably the phenomenon of superconductivity is the most spectacular example of the symmetry breaking  which a macroscopic body spontaneously undergoes. Of course this is not the only one. Antiferromagnets, liquid crystals and other states of matter obey this rather general  scheme of broken symmetries. 
The general idea is that when the number $N$ of microscopic quantum constituents which, depending on the system, corresponds to the number of Cooper pairs, particles or spins, goes to infinity, the matter undergoes a "phase" transition to a state in which the microscopic symmetries are violated \cite{and72,and52,and58,nam60,gol62}. In the context of quantum magnetism \cite{lie62,kai89,kap90}, the macroscopic classical state has been described as a combination of ``thin spectrum'' states emerging in the $N \rightarrow \infty$ limit because of the singular nature of the thermodynamic limit. The same description has been shown to apply also to the cases of quantum crystals, Bose-Einstein condensates and superconductors \cite{wez05,wez06,wez07,bir07,wez08}. 
The theory of spontaneous symmetry breaking explains the stability and rigidity of states which are not eigenstates of the underlying microscopic Hamiltonian but it makes no assertion on whether or how a symmetry broken groundstate can evolve out of the symmetric state in a closed quantum system. 
To investigate this, one can perform a {\it Gedankenexperiment} in which  a  symmetry breaking perturbation is slowly 
 switched on
in an arbitrary large but finite system initially prepared in a fully symmetric state.  
Using a particular antiferromagnetic model system -- the  Lieb-Mattis model \cite{lie62} --  it has been recently shown that the corresponding quantum dynamics  is dominated by highly non-adiabatic processes triggering the appearance of a symmetric non-equilibrium state that recursively collapses at punctured times  into a symmetry broken state \cite{ort11}. 

Here we shed light on this far-from-equilibrium symmetry-breaking process that is so ubiquitous in physics -- in the formation of crystalline matter, atomic condensates, Josephson junction arrays and local pairing superconductors to name but a few -- by introducing quantum state tomography. 
In complete analogy with medical diagnostics where three-dimensional images of the inner part of a body can be extracted from NMR or X-rays  two-dimensional images  obtained at different directions, quantum state tomography determines a quasi-probability distribution in phase space from only position ($Q$) or momentum ($\Pi$) measurements \cite{lei96,sko03}. 
This quasi probability distribution has been  introduced by Eugene Wigner in his phase space formulation of quantum mechanics. For a pure quantum state, the Wigner distribution $W(Q,\Pi)$ is defined in terms of the position wave function $\Psi(Q)$ as  $W(Q,\Pi)=\pi^{-1}  \int_{-\infty}^{\infty} \Psi^{\star} (Q-S) \Psi(Q+S) \mathrm{e}^{-2 i \Pi S} d S$ (in $\hbar=1$ units)   and retains the marginal probability distributions \cite{lei03}
$$\int W(Q,\Pi) d\Pi = |\Psi(Q)|^2, $$  
$$\int W(Q,\Pi) d Q =|\Psi(\Pi)|^2.  $$
By detecting the position of many objects prepared in the same quantum state yields the spatial distribution of the wavefunction $|\Psi(Q)|^2$ as a space-like shadow of the Wigner distribution. This, in turns, allows for a tomographic reconstruction of $W(Q,\Pi)$ once various shadows at different directions in phase space have been observed. 
Therefore, although humans are chained in a Plato's quantum cave and are entitled to see only shadows of a quantum object, they can access full quantum snapshots once these shadows are properly rotated \cite{lei98}. 

We unravel these snapshots in the dynamics of symmetry breaking by 
using the paradigmatic example of an harmonic crystal
to show that the quantum dynamics  is  generically characterized by the appearance  of revivals of the symmetric ground state wavefunction followed by collapses towards a quasi-classical state akin the symmetry broken groundstate of infinite size system.  In this quasi-classical state, however, 
the matter wave has maximum uncertainty both in total position -- precisely as in the symmetric translational invariant ground state -- and in total momentum. 
The exceptions are  punctured times which  render a Dirac comb of symmetry broken states \cite{ort11}. 
Interestingly we find this sequence of collapses/revivals of the ground state wavefunction to occur on a characteristic timescale set
by Zurek's equation of non-equilibrium quantum phase transition \cite{zur05}. 

\section{Spontaneous symmetry breaking and the thin spectrum.}
The Nambu-Goldstone theorem \cite{gol61,nam61}
 guarantees the existence of low-energy gapless collective excitations in systems with spontaneously broken continuous symmetries. 
The low-energy Hamiltonian for these normal modes, which depending on the particular system at hand correspond to phonons, spin waves or Bogoliubov's excitations can be always recast in the form
\begin{equation}
{\cal H}=\sum_{\bf k}  \, \omega_{{\bf k}} \, b^{\dagger}_{\bf k} \, b_{\bf k}, 
\end{equation}
where $\omega_{\bf k}$ indicate the frequencies of the Goldstone mode excitations vanishing in the long-wavelength limit. 
If there is a spontaneously broken symmetry, the motion along the continuous symmetry axis 
characterizing the quantum mechanics of the macroscopic body as whole
cannot be of the form \cite{bir07}
$b^{\dagger} b$   but  it will be either given by $\Pi^2/ (2 I) $ (as in crystals and Josephson junction arrays)  or $Q^2 / (2 I)$ (e.g. antiferromagnets and local pairing superconductors) \cite{wez06,wez08}
where $Q$ and  $\Pi$ are the coordinate and the conjugate  momentum operator along the symmetry axis 
whereas the parameter $I$ depends on the total number of microscopic quantum constituents and diverges in the thermodynamic limit where $N \rightarrow \infty$. 
This is explicitly manifest by considering  the specific example of an  harmonic crystal with Hamiltonian
\begin{equation}
{\cal H} = \sum_j \dfrac{p_j^2}{2m} + \dfrac{\kappa}{2} \sum_j (x_j - x_{j+1})^2,
\label{eq:hamiltoniancrystal}
\end{equation}
where the index $j$ labels the $N$ atoms in the lattice which have mass $m$, momentum  $p_j$, and position $x_j$. The harmonic potential among neighboring atoms is parametrized by the constant $\kappa$. 
 Let us now define the bosonic annihilation and creation operator
$
b_j = \left[C \,x_j + i  p_j / C \right ] /  \sqrt{2\hbar}$ and  $ b^{\dagger}_j=\left[C \,x_j - i  p_j / C \right ] / \sqrt{2\hbar}$
where $C=(2 \,m \, \kappa)^{1/4}$. In momentum space, the Hamiltonian Eq.(\ref{eq:hamiltoniancrystal}) reduces to
$$
{\cal H} = \hbar \sqrt{\dfrac{\kappa}{2m}}\sum_k \left (A_k b_k^{\dagger}b_k + \dfrac{B_k}{2}(b_k^{\dagger}b_{-k}^{\dagger} + b_kb_{-k}) + 1 \right ),
$$
where $A_k = 2 - \cos(ka)$, $B_k = -\cos(ka)$ and $a$ is the lattice constant.
This Hamiltonian is not diagonal since the terms $b_k^{\dagger}b_{-k}^{\dagger}$ and $b_kb_{-k}$ create and annihilate two bosons at the same time. One can get rid of them by performing a canonical Bogoliubov transformation. However, the parameters in the Bogoliubov transformation diverge as $k \rightarrow 0$ and thus the canonical transformation is not well defined \cite{wez06}. This implies that one should investigate separately the $k=0$ component
$${\cal H}_{k=0}= \hbar \sqrt{\dfrac{\kappa}{2m}} \left[1-\dfrac{1}{2} \left(b^{\dagger}_0-b_0\right)^2 \right].$$
This part of the Hamiltonian describes the fact that the quantum crystal carries a kinetic energy associated with the combined mass of all $N$ atoms. Going back to real space it reads
\begin{equation}
{\cal H}_{k=0}=\dfrac{\Pi^2}{2 \, m \, N}, 
\label{eq:hamiltoniancrystalk0}
\end{equation}
where $\Pi=\sum_j p_j$ is the total momentum of the entire system and we left out a negligible constant. 
The ground state of this Hamiltonian
has total zero momentum: it has no uncertainty in total momentum and maximum uncertainty in total position thereby implying that the translational symmetry is unbroken. 
At finite $N$, the excitations over the ground state respecting the symmetry give rise to a tower of ultra-low energy states that form the so-called ``thin spectrum''\cite{wez06,wez07}. 
It is "thin" because it contains states that are of such low energy that their contribution to thermodynamic quantities vanish in the thermodynamic limit. Nevertheless, when $N \rightarrow \infty$, the thin spectrum excitations collapse to form a degenerate continuum of states. Within this continuum even a vanishingly small symmetry breaking perturbation is enough to couple different thin spectrum states thereby stabilizing the symmetry broken ground state. 
To show this, let us take into account  the effect of a symmetry breaking perturbation -- a pinning potential for the individual atoms-- rendering a symmetry breaking Hamiltonian ${\cal H}_{SB}=- B \sum_{j}  \cos\left(2 \pi x_j \right) / (2 \pi)^2$.  For small deviations of the atoms from their mean positions, the zero momentum term in lowest order is given \cite{wez06} by ${\cal H}_{SB}= B x_{tot}^2 / (2 N)$ where $x_{tot}=\sum_j x_j$. 
This, in turn, implies  that the collective behaviour of the harmonic crystal as a whole is governed by the harmonic-oscillator Hamiltonian 
\begin{equation}
{\cal H}_{c}=\dfrac{\Pi^2}{2 \, N} + \dfrac{B \, N \, Q^2}{2},
\end{equation}
where for simplicity we have set $m=1$ and we introduced the centre of mass coordinate $Q=x_{tot} / N$  satisfying the canonical commutation relation $\left[Q, \Pi \right] = i \hbar$. 
The quantum of energy of this Hamiltonian $\propto \sqrt{B}$ and the excitations over the ground state 
realize a "dual" thin spectrum. The ground state wavefunction corresponds to a Gaussian wave packet for the collective coordinate $Q$ of the form $\Psi_0 (Q) \propto \mathrm{e}^{-Q^2 / 2 L^2}$ with width $L \propto \left( N \, \sqrt{B} \right)^{ - 1 / 2}$. 
For a vanishing symmetry breaking field and a finite number of atoms, we find that the ground state wavefunction obviously collapses onto the symmetric ground state of the microscopic Hamiltonian. However, by taking first the thermodynamic limit 
the centre of mass position becomes completely localised even if at the end 
 the symmetry breaking field is sent to zero.  Therefore one finds that the system is in a stable state which is not an eigenstate of the underlying microscopic Hamiltonian -- the system is inferred to spontaneously break the symmetry. 
Strictly speaking, only truly infinite size systems are allowed to spontaneously break the symmetry. A large, but not infinitely large crystal requires a finite symmetry breaking field to stabilize a symmetry broken state over the symmetric ground state of the microscopic Hamiltonian. 

\begin{figure}
\includegraphics[width=.8\columnwidth]{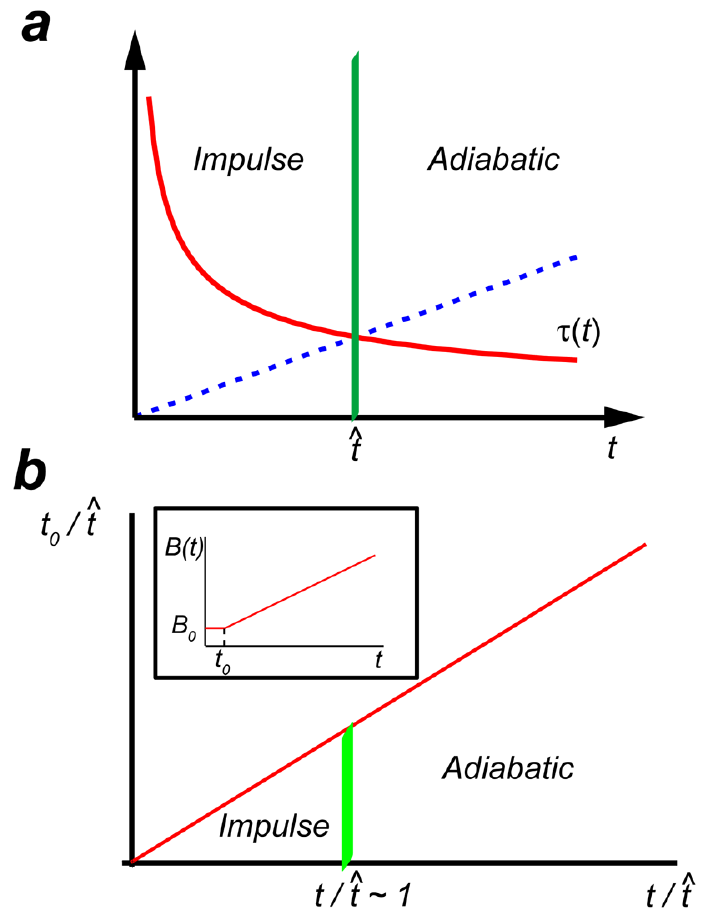}
\caption{ (Color online). (a) Sketch of the behavior of the relaxation time (continuous line)  as compared to the typical time on which the Hamiltonian is changed  $t$ determining the freeze-out time of  Zurek's equation. (b) The corresponding regimes for the dynamics of symmetry breaking in the $t-t_0$ plane. Times have been measured in units of the freeze-out time $\hat{t}$. The inset shows the setup of the symmetry breaking field.}
\label{fig:phasediagram}
\end{figure}

\section{Quantum dynamics of symmetry breaking} 
Let us then consider such a large but finite system  with a pinning potential whose strength is switched on linearly in time as $B(t)=\delta t$ with ramp rate $\delta$. At initial time $t_0$ we consider a finite symmetry breaking perturbation $B(t_0) \equiv B_0$ [c.f. inset of Fig.~\ref{fig:phasediagram}(b)]  and the wave function of the system in the corresponding ground state. We stress that the choice of a finite symmetry breaking perturbation at initial time is  essential in order to have a cutoff guaranteeing the continuity of the wave function basis. Lateron we will consider the  limit $B_0 \rightarrow 0$ corresponding for  finite $N$ to an initial completely symmetric ground state with gapless dual thin spectrum excitations.
The inclusion of a cutoff $B_0$ renders  two distinct regimes of the quantum dynamics. Whenever the characteristic relaxation time $\tau=\left(\delta t\right)^{-1/2}$ is much smaller than 
the typical timescale $t$ on which the Hamiltonian is changed, the system is able to react to the changing Hamiltonian thereby rendering an adiabatic passage. In the opposite regime  $\tau  \gg t$,  reflexes of the system are so deteriorated that the state can be considered effectively frozen and the dynamics is impulse-like.  Clearly, the crux of this story is the instant where the dynamics changes from impulse to adiabatic --it is determined by Zurek's equation \cite{zur05} $\tau(t) \equiv t $ and defines the so-called freeze-out time $\hat{t} \propto \delta^{-1/3}$ [c.f. Fig.~\ref{fig:phasediagram}(a)]. By considering an initial time $t_0 \gg \hat{t}$, the entire evolution will be thus nearly  adiabatic. 
In this case, fluctuations of the centre of mass coordinate decreases continuously in time as $\Delta Q^2 =\langle Q^2 \rangle - \langle Q \rangle ^2 \propto \left[ N \sqrt{\delta t} \right]^{-1} $.  
However, for the dynamics of symmetry breaking to be adiabatic the ramp rate $\delta$ is seen to be bounded by $\delta < B_0^{3/2} $. 
Henceforth for a vanishing ramp rate $\delta$ at finite values of $B_0$, we recover a quasi-adiabatic time evolution. But taking the $B_0 \rightarrow 0$ limit at finite ramp rate $\delta$, the dynamics will start in the impulse regime even if at the end the ramp rate is sent to zero -- the adiabatic limit can never be reached for a sufficiently small $B_0$.
This is in agreement with the recent finding \cite{pol08} that adiabatic processes in low-dimensional systems with broken continuous symmetries are absent. 

\begin{center}
\begin{figure*}
\includegraphics[width=\textwidth]{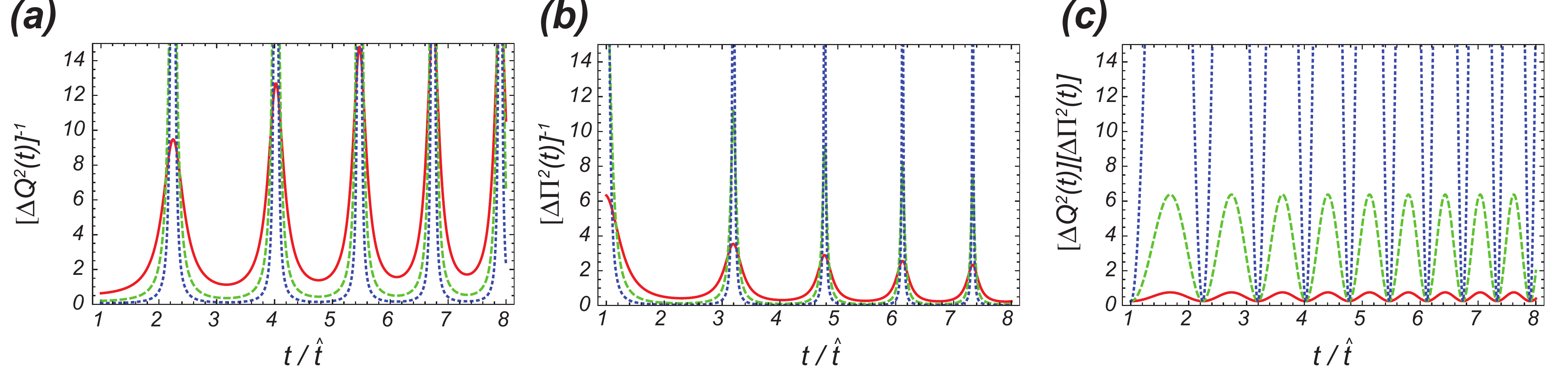}
\caption{ 
(Color online) (a) Time evolution of $(\Delta Q^{2})^{-1}$ for different values of the initial time $t_0$. The continuous thick line is for $t_0=10^{-1}$, the dashed line for $t_0=10^{-2}$ and the dotted line corresponds to $t_0=10^{-3}$. By decreasing the initial time $t_0$, a Dirac comb of symmetry-broken states is approached. (b) Same for the inverse of the fluctuations of the total momentum. Also in this case a Dirac comb is approached.  (c) Same of (a) for the time evolution of the total uncertainty of the matter wave. Apart from punctured times, it is seen to diverge in the $t_0 \rightarrow 0$ limit. }
\label{fig:fig1}
\end{figure*}
\end{center}
In the same spirit of Ref.~\onlinecite{ort11}, to analyse the dynamics of symmetry breaking in the strongly non-adiabatic regime, we first make use of the adiabatic-impulse (AI) approximation which underlies the Kibble-Zurek theory \cite{kib76,zur85,zur96} of non-equilibrium phase transition. 
In the AI scheme \cite{dam06}, the initial state is considered effectively frozen in the impulse-like regime $t_0 < t < \hat{t}$ changing only by a trivial overall phase factor. 
At freeze-out time, the system therefore reaches a state that is a superposition of dual thin spectrum excitations  $\ket{\Psi_0(Q, t_0)} = \sum_n c_n \ket{\Psi_n (Q, \hat{t})}$ 
where the coefficients $c_n$ are non zero only for even values of the quantum number $n$. 
The evolution at $t>\hat{t}$ can be considered to be adiabatic and therefore the dynamics of the wavefunction 
is governed  by $\ket{\Psi (Q, t)}=\sum_n \, c_n\ket{\Psi_n(Q, t)}e^{-i\Omega_n(t)}$
where we have defined the dynamical phase factor
$\Omega_n (t)=2 / 3 \times  \left[\left(t / \hat{t}\right)^{3/2}-1\right] \left(n+ 1/2\right). $
Within the AI approximation, we can obtain the  time-evolution of the symmetry breaking order parameter -- defined by the inverse of the fluctuations of the centre of mass coordinate-- taking explicitly into account quantum phase interference effects. This is unlike for instance the case of the Lieb-Mattis model where in the AI scheme
the time evolution of the  staggered magnetisation can be computed only by neglecting interference effects \cite{ort11}. 
We find
\begin{equation}
\left[\Delta Q^2 (t)\right]^{-1} =2 \dfrac{N}{\hat{t}} \sqrt{\dfrac{t\,t_0}{\hat{t}^2}}\left[1-\left(1-\frac{t_0}{\hat{t}}\right)\sin^2\Omega(t)\right]^{-1},
\end{equation}
the behaviour of which  is shown in Fig.~\ref{fig:fig1}(a). When decreasing the initial time $t_0$ we observe that the behaviour of the order parameter calculated above corresponds precisely to a Dirac comb of symmetry broken states in perfect agreement with the case of antiferromagnets \cite{ort11}. 
Besides this, we find that  at the punctured times where the fluctuation of the centre of mass coordinate vanishes the dynamical phases of the excited dual thin spectrum excitations have $\pi$ shifts, {\it i.e.} for $t= t_k = \left[3 \kappa \pi / 2+ 3 \pi / 4 + 1 \right]^{2/3}  $ with $\kappa$ integer.   
This shows that independent of the actual strength of the symmetry breaking perturbation,  destructive quantum phase interference leads to an  instantaneous breaking of the translational symmetry. It is also manifested by the fact that a direct computation shows  at these instants ${\hat Q} \ket{\Psi (Q, t_k)} \equiv 0$ and therefore the harmonic crystal is completely localised in the centre of the potential well. 

To further show the nature of the non-equilibrium state realised in the remaining time evolution, we have determined the time-evolution of the inverse of the fluctuations of the total momentum of the entire crystal and find
\begin{equation}
\left[\Delta\Pi^2(t)\right]^{-1}= 2 \frac{\hat{t}}{N}\sqrt{\frac{\hat{t}^2}{t\,t_0} }\left[1-\left(1-\frac{\hat{t}}{t_0}\right)\sin^2\Omega(t)\right]^{-1},
\end{equation}
the behaviour of which is shown in Fig.~\ref{fig:fig1}(b).   
A Dirac comb is also the result.
 The instants where the system is  an eigenstate of the total momentum $\Pi$ -- which correspond to  revivals of the initial completely delocalised symmetric state even in the presence of a sizable  pinning potential --  are realised for $t=t_k= \left[3 \kappa \pi / 2+  1 \right]^{2/3}$ 
 with $\kappa$ integer in which case
 quantum  phase interference effects are absent. 
This is again in line with the dynamics of the  Lieb-Mattis model and henceforth guarantees the universality of the dynamical symmetry breaking phenomenon independent of the specific microscopic model taken into account.

Apart form the punctured times where the matter wave has either no uncertainty in total position or no uncertainty in total momentum, a strongly non-equilibrium state is realised. 
This state is rather interesting as it retains a perfect delocalisation of both the centre of mass position and its corresponding momentum. It has an infinite uncertainty, {\it i.e.} $\Delta\Pi^2 \Delta Q^2 \rightarrow \infty$ as it is shown in Fig.~\ref{fig:fig1}c.

\section{Quantum dynamics in phase space} To 
unravel the origin of 
 this non-equilibrium state
 we analyse the dynamics of symmetry breaking with quantum state tomography.  
The time-dependent Hamiltonian
\begin{equation}
{\cal H}(t)= \dfrac{\Pi^2}{2  \, N}+\dfrac{N}{2} \, \delta \, t \, Q^2 , \nonumber 
\label{eq:hamiltoniantimedependent}
\end{equation}
represents a  simple example of generalized time-dependent harmonic oscillator whose exact quantum theory has been extensively studied in the literature \cite{kha86,kha79,col81,col82,col83,col85,abd86,abda86,pap74,yeo93,lim08,son99,son00}. 
In particular, within the Feynman path integral approach it has been shown \cite{son99} that the spectral decomposition of the propagator ${\cal G}(Q_b,t_b | Q_a,t_a)=\sum_n \Psi^{\star}_n(Q_a,t_a) \, \Psi_n(Q_b,t_b)$ is defined by a complete set of wavefunctions of the form
\begin{eqnarray}
\Psi_{n}(Q,t)&=&\sqrt{\dfrac{1}{2^{n} \, n!}} \left[\dfrac{{\it Re}\left[ \omega(t)\right]}{\pi}\right]^{1/4} \,\,   H_{n}\left[\sqrt{{\it Re} \left[\omega(t)\right]}\, Q\right]\,\nonumber \\ & & 
 \mathrm{e}^{-\frac{ Q^2}{2} \omega(t)} \times \mathrm{e}^{-\mathrm{i} \left(n+\frac{1}{2}\right) \phi(t)} \, 
\label{eq:dynamicwavefunction}
\end{eqnarray}
\begin{figure}
\includegraphics[width=\columnwidth]{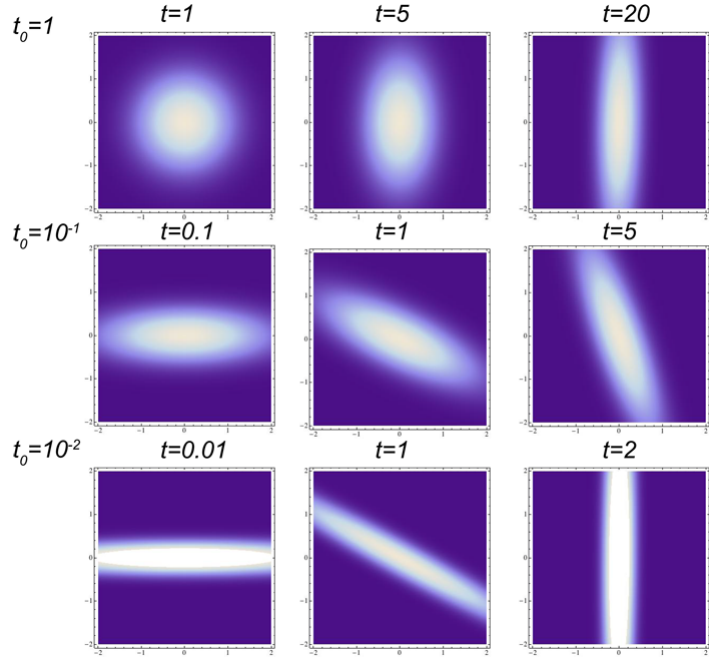}
\caption{(Color online) Density plots of the Wigner function in phase space $(Q, \Pi)$. The centre of mass coordinate $Q$ and the total momentum $\Pi$ have been rescaled  by $N^{-1/2}$ and $N^{1/2}$ respectively in order to absorb the $N$ dependence of the Wigner function. By decreasing the initial time $t_0$ the time evolution changes from an adiabatic shear motion, to a strongly non-equilibrium rotative motion.}
\label{fig:fig3}
\end{figure}
where $H_n$ are the Hermite polynomials and ${\it Re}\left[\omega(t)\right]>0$ in order to guarantee square integrability. 
The quantal phase $\phi(t)$ and the complex parameter $\omega(t)$ are uniquely determined by solving the classical Euler-Lagrange equation of motion. 
Different sets of wave function of the form Eq.~\ref{eq:dynamicwavefunction} correspond to take different pairs of linearly independent solutions to the classical equation of motion. 
This enables to choose two particular solutions guaranteeing that at the initial time  $\Psi_0(Q,t_0)$ corresponds to the initial static Gaussian wavepacket and  implies  that the wavefunction at all times remains an $n=0$ state of the form of Eq.~\ref{eq:dynamicwavefunction}.

\begin{figure}
\includegraphics[width=.9\columnwidth]{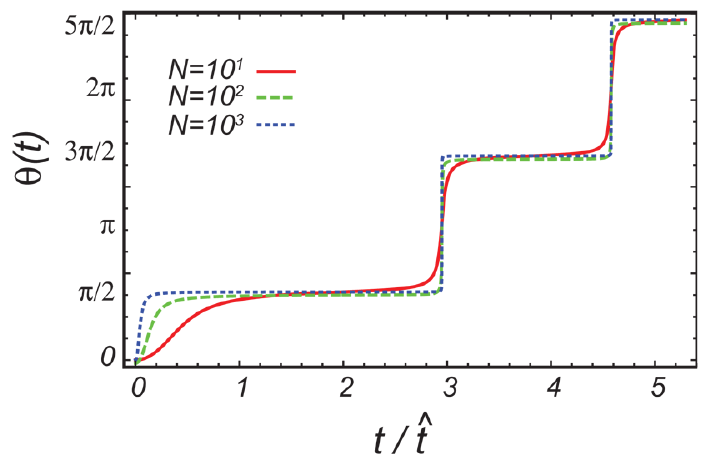}
\caption{(Color online) The time evolution of the angle $\theta$ characterising the rotation in phase space of the Wigner function. By increasing the number of microscopic constituents, a step-like behaviour is realised. }
\label{fig:wignerbis}
\end{figure}

The exact solution of the 
 time-dependent Schr\"odinger equation 
allows us to determine the time evolution of the Wigner function [see Fig.\ref{fig:fig3}] given by 
\begin{equation}
W(Q,\Pi,t)=\dfrac{1}{\pi} \mathrm{e}^{-{\it Re}[\Omega(t)] Q^2} \times \mathrm{e}^{-\left(\Pi + {\it Im}[\Omega(t)]  Q \right)^2 / {\it Re}[\Omega(t)]}. 
\label{eq:wigner}
\end{equation}
By considering a nearly-adiabatic process --  an initial time $t_0 >> \hat{t}$ -- one finds that the  Wigner function shears in time 
in agreement with the time evolution of a Gaussian wavepacket reacting adiabatically to the time change of the harmonic oscillator angular frequency. 
On the contrary,  
in the out-of-equilibrium regime, {\it i.e.} for $t_0 < \hat{t}$,
the Wigner distribution shears and rotates in phase space as it follows from the fact that the initial width of the Gaussian wavepacket acquires a non-negligible imaginary part. Finally, in the $t_0 \rightarrow 0$ limit instead, any shear is absent and the motion simply corresponds to a rigid rotation in phase space. 
For a completely symmetric initial state,  the initial Wigner function corresponds to an infinite line in phase space $\delta(Q)$ and a rigid rotation is the only motion preserving this one-dimensional character. 
As a result, we find in the $t_0 \rightarrow 0$ limit the time evolution of the Wigner function as
\begin{equation}
 W(Q,\Pi,t) \simeq \delta( \cos{\theta(t)} \, Q + \sin{\theta(t)} \, \Pi),
 \label{eq:wigner2}
 \end{equation}
where the angle $\theta(t) \simeq \tan^{-1} {\it Im}[\Omega(t)] $.  By increasing the number of microscopic quantum constituents $N$, the time dependence of  the angle $\theta(t)$ approaches a step-like behavior [c.f. Fig.~\ref{fig:wignerbis}] thereby implying that the quantum dynamics of symmetry breaking in a macroscopic body is characterized by revivals of the initial symmetric state  and 
collapses  of the initial quantum state towards a "quasi-classical" state -- tomographically indistinguishable from the  symmetry broken state of infinite size systems but, as we have shown above, completely delocalised both in momentum and in real space.

\section{Conclusions}
\label{sec:conclusions}
In conclusion, by considering the paradigmatic example of an harmonic oscillator, we have shown that in the dynamical realm symmetry breaking is characterized by far-from-equilibrium processes. 
No matter how slowly a symmetry breaking perturbation is driven, the adiabatic limit can never be reached in a macroscopic body. 
By means of quantum state tomography, we have shown that 
nevertheless
the evolution of symmetry breaking corresponds to a continuous, rigid rotation of the Wigner distribution. This rotation yields at the same time a sequence of 
step-like revivals of the symmetric state followed by 
collapses onto a
symmetry broken groundstate akin to the one realized in infinite size systems but with maximum uncertainty both in total position and in total momentum.

\section{Acknowledgements} 
We  thank P. Marra for valuable discussions.


\begin{thebibliography}{10}

\bibitem{and72}
P.~W. Anderson, Science {\bf 177},  393  (1972).

\bibitem{and52}
P.~W. Anderson, Phys. Rev. {\bf 86},  694  (1952).

\bibitem{and58}
P.~W. Anderson, Phys. Rev. {\bf 112},  1900  (1958).

\bibitem{nam60}
Y. Nambu, Phys. Rev. {\bf 117},  648  (1960).

\bibitem{gol62}
J. Goldstone, A. Salam, and S. Weinberg, Phys. Rev. {\bf 127},  965  (1962).

\bibitem{lie62}
E. Lieb and D. Mattis, Journal of Mathematical Physics {\bf 3},  749  (1962).

\bibitem{kai89}
C. Kaiser and I. Peschel, Journal of Physics A: Mathematical and General {\bf
  22},  4257  (1989).

\bibitem{kap90}
T.~A. Kaplan, W. von~der Linden, and P. Horsch, Phys. Rev. B {\bf 42},  4663
  (1990).

\bibitem{wez05}
J. van Wezel, J. van~den Brink, and J. Zaanen, Phys. Rev. Lett. {\bf 94},
  230401  (2005).

\bibitem{wez06}
J. van Wezel, J. Zaanen, and J. van~den Brink, Phys. Rev. B {\bf 74},  094430
  (2006).

\bibitem{wez07}
J. van Wezel and J. van~den Brink, American Journal of Physics {\bf 75},  635
  (2007).

\bibitem{bir07}
T. Birol, T. Dereli, O.~E. M\"ustecapl\ifmmode \imath \else \i
  \fi{}o\ifmmode~\breve{g}\else \u{g}\fi{}lu, and L. You, Phys. Rev. A {\bf
  76},  043616  (2007).

\bibitem{wez08}
J. van Wezel and J. van~den Brink, Phys. Rev. B {\bf 77},  064523  (2008).

\bibitem{ort11}
C. Ortix, J. Rijnbeek, and J. {van den Brink}, Phys. Rev. B {\bf 84},  144423
  (2011).

\bibitem{lei96}
D. Leibfried, D. Meekhof, B. King, C. Monroe, W. Itano, and D. Wineland, Phys.
  Rev. Lett. {\bf 77},  4281  (1996).

\bibitem{sko03}
E. Skovsen, H. Stapelfeldt, S. Juhl, and K. Molmer, Phys. Rev. Lett. {\bf 91},
  090406  (2003).

\bibitem{lei03}
D. Leibfried, R. Blatt, C. Monroe, and D. Wineland, Rev. Mod. Phys. {\bf 75},
  281  (2003).

\bibitem{lei98}
D. Leibfried, T. Pfau, and M. Christopher, Physics Today  (1998).

\bibitem{zur05}
W.~H. Zurek, U. Dorner, and P. Zoller, Phys. Rev. Lett. {\bf 95},  105701
  (2005).

\bibitem{gol61}
J. Goldstone, Nuovo Cimento {\bf 19},    (1961).

\bibitem{nam61}
Y. Nambu and G. {Jona Lasinio}, Phys. Rev. {\bf 122},    (1961).

\bibitem{pol08}
A. Polkovnikov and V. Gritsev, Nat Phys {\bf 4},  477  (2008).

\bibitem{kib76}
T.~W.~B. Kibble, Journal of Physics A: Mathematical and General {\bf 9},  1387
  (1976).

\bibitem{zur85}
W.~H. Zurek, Nature (London) {\bf 317},  505  (1985).

\bibitem{zur96}
W.~H. Zurek, Physics Reports {\bf 276},  177  (1996).

\bibitem{dam06}
B. Damski and W.~H. Zurek, Phys. Rev. A {\bf 73},  063405  (2006).

\bibitem{kha86}
D.~C. {Khandekar} and S.~V. {Lawande}, Phys. \ Rep. {\bf 137},  115  (1986).

\bibitem{kha79}
D.~C. {Khandekar} and S.~V. {Lawande}, Journal of Mathematical Physics {\bf
  20},  1870  (1979).

\bibitem{col81}
R. {Colegrave} and M. {Abdalla}, J. Phys. A. {\bf 14},  2269  (1981).

\bibitem{col82}
R. {Colegrave} and M. {Abdalla}, J. Phys. A. {\bf 15},  1549  (1982).

\bibitem{col83}
R. {Colegrave} and M. {Abdalla}, J. Phys. A. {\bf 16},  3805  (1983).

\bibitem{col85}
M. {Sebawe Abdalla} and R.~K. {Colegrave}, Phys. \ Rev. \ A {\bf 32},  1958
  (1985).

\bibitem{abd86}
M.~S. {Abdalla}, Phys. \ Rev .\ A {\bf 34},  4598  (1986).

\bibitem{abda86}
M.~S. {Abdalla}, Phys. \ Rev .\ A {\bf 33},  2870  (1986).

\bibitem{pap74}
G.~J. {Papadopoulos}, Journal of Physics A Mathematical General {\bf 7},  209
  (1974).

\bibitem{yeo93}
K.~H. {Yeon}, K.~K. {Lee}, C.~I. {Um}, T.~F. {George}, and L.~N. {Pandey},
  Phys. \ Rev. \ A {\bf 48},  2716  (1993).

\bibitem{lim08}
A. {Lopes de Lima}, A. {Rosas}, and I.~A. {Pedrosa}, Annals of Physics {\bf
  323},  2253  (2008).

\bibitem{son99}
D.-Y. Song, Phys. Rev. A {\bf 59},  2616  (1999).

\bibitem{son00}
D.-Y. Song, Phys. Rev. Lett. {\bf 85},  1141  (2000).

\end{thebibliography}
\end{document}